\documentclass[pra,twocolumn,floatfix,showpacs,superscriptaddress,nofootinbib]{revtex4}

\usepackage{times,amsmath,amsfonts,amssymb,latexsym}
\usepackage{graphicx}
\usepackage{epsfig}

\newenvironment{proof}[1][Proof]{\noindent\textbf{#1.} }{\ \rule{0.5em}{0.5em}}

\newtheorem{myproposition}{Proposition}

\newtheorem{mydefinition}{Definition}
\newtheorem{mytheorem}{Theorem}

\newtheorem{mycorollary}{Corollary}

\begin{document}

\title{Maps for general open quantum systems and a theory of linear quantum
error correction}
\author{Alireza Shabani}
\affiliation{Department of Electrical Engineering-Systems, University of Southern
California, Los Angeles, CA 90089, USA}
\author{Daniel A. Lidar}
\affiliation{Department of Electrical Engineering-Systems, University of Southern
California, Los Angeles, CA 90089, USA}
\affiliation{Departments of Chemistry and Physics, University of Southern California, Los
Angeles, CA 90089, USA}

\begin{abstract}
We show that quantum subdynamics of an open quantum system can always be
described by a Hermitian map, irrespective of the form of the initial total
system state. Since the theory of quantum error correction was developed
based on the assumption of completely positive (CP)\ maps, we present a
generalized theory of linear quantum error correction, which applies to any
linear map describing the open system evolution. In the physically relevant
setting of Hermitian maps, we show that the CP-map based version of quantum
error correction theory applies without modifications. However, we show that
a more general scenario is also possible, where the recovery map is
Hermitian but not CP. Since non-CP maps have non-positive matrices in their
range, we provide a geometric characterization of the positivity domain of
general linear maps. In particular, we show that this domain is convex, and
that this implies a simple algorithm for finding its boundary.
\end{abstract}

\maketitle

\section{Introduction}

The problem of the formulation and characterization of the dynamics of
quantum open systems has a long and extensive history \cite%
{Davies:76,Alicki:87,Breuer:Book}. This problem has become particularly
relevant in the context of quantum information processing \cite{Nielsen:book}%
, where a remarkable theory of quantum error correction (QEC) was developed
in recent years to address the problem of how to process quantum information
in the presence of decoherence and imperfect control \cite{Gaitan:book}. A\
key assumption common to many previous QEC studies is that the evolution of
the quantum information processor can be described by a succession of \emph{%
completely positive}\ (CP) maps \cite{Kraus:83}, interrupted by unitary
gates or measurements \cite{Knill:97b}. However, it is well known that if
the initial total system state is entangled, quantum dynamics is not
described by a CP\ map \cite%
{Pechukas+Alicki:95,Stelmachovic:01,Jordan:04,Carteret:05,Rodriguez:07}. In
fact, we showed very recently in Ref. \cite{ShabaniLidar:08} that a CP\ map
arises if and only if the initial total system state has vanishing quantum
discord \cite{Ollivier:01}, i.e., is purely classically correlated. One is
thus naturally led to ask whether this impacts the applicability of QEC\
theory under circumstances where non-classical initial state correlations
play a role. Here \textquotedblleft initial state\textquotedblright\ does
not refer exclusively to the \textquotedblleft $t=0$\textquotedblright\
point, but also to intermediate times where the recovery map is applied,
since this map was also assumed to be CP in standard quantum error
correction theory \cite{Knill:97b}. Motivated by this fact we here
critically revisit the CP maps assumption in QEC, and show that it can be
relaxed \footnote{%
Note that this is issue is entirely distinct from the critique of Markovian
fault tolerant QEC\ expressed in \cite{AlickiLidarZanardi:05}, which was
concerned with the compatibility of other assumptions of fault-tolerant QEC
(specifically, fast gates and pure ancillas) with rigorous derivations of
the Markovian limit.}. To do so, we first consider the problem of
characterizing the type of map that describes open system evolution given an 
\emph{arbitrary} initial total system state (Section \ref{sec:QDP}). We show
that this map is always a linear, \emph{Hermitian} map (of which CP\ maps
are a special case). We then argue that the generic noise map describing the
evolution of a quantum computer as it undergoes fault tolerant quantum error
correction (FT-QEC)\ is indeed not a CP map, but rather such a Hermitian,
linear map (Section \ref{sec:FTQEC}). The reason is, essentially, that
imperfect error correction results in residual non-classical correlations
between the system and the bath, as the next QEC\ cycle is applied. To deal
with this, we develop a generalized theory of QEC which we call
\textquotedblleft linear quantum error correction\textquotedblright\ (LQEC),
which applies to arbitrary linear maps on the system (Section \ref{sec:LQEC}%
). Then we show that, fortunately, the CP-map based version of QEC theory
applies without modifications in the physically relevant setting of
Hermitian maps. However, we show that a more general scenario is also
possible, where the recovery map is Hermitian but not CP. This is useful
since it obviates the unrealistic assumption that the recovery ancillas
enter the QEC cycle as classically correlated with the other system qubits.
Our results significantly extend the realm of applicability of QEC, in
particular to arbitrarily correlated system-environment states. We conclude
in Section \ref{sec:conc}.

\section{Quantum dynamical processes and maps}

\label{sec:QDP}

In this section we prove a basic new result, that a quantum dynamical
process can always be represented as a linear, Hermitian map from the
initial to the final system-only state. In doing so we rely heavily on our
previous work \cite{ShabaniLidar:08}.

The dynamics of open quantum systems can be described as follows. Consider a
quantum system $S$ coupled to another system $B$, with respective Hilbert
spaces $\mathcal{H}_{S}$ and $\mathcal{H}_{B}$, such that together they form
one isolated system, described by the joint initial state (density matrix) $%
\rho _{SB}(0)$. Their joint time-evolved state is then 
\begin{equation}
\rho _{SB}(t)=U(t)\rho _{SB}(0)U^{\dag }(t),
\end{equation}%
where $U(t)$ is the unitary propagator of the joint system-bath dynamics
from the initial time $t=0$ to the final time $t$, i.e., the solution to the
Schrodinger equation $\dot{U}=-(i/\hbar )[H,U]$, where $H$ is the joint
system-bath Hamiltonian. The object of interest is the system $S$, whose
state at all times $t$ is governed according to the standard
quantum-mechanical prescription by the following quantum dynamical process
(QDP): 
\begin{equation}
\rho _{S}(t)=\mathrm{Tr}_{B}[\rho _{SB}(t)]=\mathrm{Tr}_{B}[U_{SB}(t)\rho
_{SB}(0)U_{SB}(t)^{\dag }].  \label{dynamics1}
\end{equation}%
$\mathrm{Tr}_{B}$ represents the partial trace operation, corresponding to
an averaging over the bath degrees of freedom \cite{Breuer:Book}.

The QDP (\ref{dynamics1}) is a transformation from $\rho _{SB}(0)$ to $\rho
_{S}(t)$. However, since we are not interested in the state of the bath, it
is natural to ask:

\begin{quotation}
Under which conditions on $\rho _{SB}(0)$ is the QDP a map $\Phi _{\mathrm{Q}%
}(t)$, 
\begin{equation}
\rho _{S}(t)=\Phi _{\mathrm{Q}}(t)[\rho _{S}(0)],  \label{eq:Qmap}
\end{equation}%
and what are the properties of this map?
\end{quotation}

In general, a\ map is an association of elements in the range with 
elements in the domain. Here we use the term \textquotedblleft
map\textquotedblright\ solely to indicate a \emph{state-independent}
transformation between two copies of the \emph{same} Hilbert space, in
particular $\mathcal{H}_{S}\mapsto \mathcal{H}_{S}$.\footnote{%
This is meant to exclude claims that system state-dependent transformations
qualify as CP\ maps, as in Ref. \cite{Tong:04}. In such cases the elements
of the transformation (the \textquotedblleft Kraus
operators\textquotedblright ) depend on the system input state, which
contradicts our notion of a map.} Then, a well-known partial answer is that
if $\rho _{SB}(0)$ is a tensor product state, i.e., $\rho _{SB}(0)=\rho
_{S}(0)\otimes \rho _{B}(0)$, then the QDP (\ref{dynamics1}) is a CP\ map. A
more general answer was provided in \cite{ShabaniLidar:08}. To explain this
answer we must first introduce some terminology.

\subsection{Various linear maps}

A map $\Phi :\mathcal{B}(\mathcal{H})\mapsto \mathcal{B}(\mathcal{H})$
[space of bounded operators on $\mathcal{H}$] is linear if $\Phi \lbrack
a\rho _{1}+b\rho _{2}]=a\Phi \lbrack \rho _{1}]+b\Phi \lbrack \rho _{2}]$
for any pair of states $\rho _{1},\rho _{2}:\mathcal{H}\mapsto \mathcal{H}$,
and constants $a,b\in \mathbb{C}$. A linear map is called Hermitian if it
maps all Hermitian operators in its domain to Hermitian operators. We first
present an operator sum representation for arbitrary and Hermitian linear
maps, that generalizes the standard Kraus representation for CP maps \cite%
{Kraus:83}. The proof is presented in Appendix \ref{app:th1}.

\begin{mytheorem}
\label{th1}A map $\Phi _{\mathrm{L}}:\mathfrak{M}_{n}\mapsto \mathfrak{M}%
_{m} $ (where $\mathfrak{M}_{n}$ is the space of $n\times n$ matrices) is
linear iff it can be represented as 
\begin{equation}
\Phi _{\mathrm{L}}(\rho )\text{ }=\sum\nolimits_{\alpha }E_{\alpha }\rho
E_{\alpha }^{\prime \dagger }  \label{eq:LM}
\end{equation}%
where the \textquotedblleft left and right operation
elements\textquotedblright\ $\{E_{\alpha }\}$ and $\{E_{\alpha }^{\prime }\}$
are, respectively, $m\times n$ and $n\times m$ matrices.\newline
$\Phi _{\mathrm{H}}$ is a Hermitian map iff 
\begin{equation}
\Phi _{\mathrm{H}}(\rho )=\sum\nolimits_{\alpha }c_{\alpha }E_{\alpha }\rho
E_{\alpha }^{\dagger },\quad c_{\alpha }\in \mathbb{R}.  \label{eq:QM}
\end{equation}
\end{mytheorem}

We will sometimes denote a linear map by listing its elements, as in $\Phi _{%
\mathrm{L}}=\{E_{\alpha },E_{\alpha }^{\prime }\}_{\alpha =1}^{r}$. Note
that a linear map $\Phi _{\mathrm{L}}=\{E_{\alpha },E_{\alpha }^{\prime
}\}_{\alpha =1}^{r}$ is trace preserving if ${\sum_{\alpha =1}^{r}}E_{\alpha
}^{\prime \dagger }E_{\alpha }=I$. Also note that the two sets of operation
elements $\{E_{\alpha },E_{\alpha }^{\prime }\}_{i=1}^{r}$ and $\{F_{\beta
},F_{\beta }^{\prime }\}_{\beta =1}^{r}$, where $F_{\beta }=\sum_{\alpha
=1}^{r}u_{\alpha \beta }E_{\alpha }$ and $F_{\beta }^{\prime }=\sum_{\alpha
=1}^{r}v_{\alpha \beta }E_{\alpha }^{\prime }$, represent the same linear
map $\Phi _{\mathrm{L}}$ if the matrices $u$ and $v$ satisfy $uv^{\dag }=I$.

As a simple example of a non-CP, Hermitian\ map, consider the\textit{\ }%
\emph{inverse-phase-flip map}. The well-known CP phase-flip map is \cite%
{Nielsen:book}: $\Phi _{\mathrm{PF}}(\rho )=\mathcal{(}1-p)\rho +p\sigma
_{z}\rho \sigma _{z}$, where $0\leq p\leq 1$ and $\sigma _{z}$ is a Pauli
matrix. Solving for $\Phi _{\mathrm{PF}}^{-1}$ from $\Phi _{\mathrm{PF}%
}^{-1}[\Phi _{\mathrm{PF}}(\rho )]=\rho $, we find that $\Phi _{\mathrm{PF}%
}^{-1}(\rho )=c_{0}\rho +c_{1}\sigma _{z}\rho \sigma _{z}$, where $%
c_{1}=p/(2p-1)$ and $c_{0}=1-c_{1}$, and $c_{0},c_{1}$ have \emph{opposite}
sign for $0<p<1$. Moreover, $\mathrm{Tr}[\Phi _{\mathrm{PF}}^{-1}(\rho )]=%
\mathrm{Tr}(\rho )$. Therefore $\Phi _{\mathrm{PF}}^{-1}$ is a
trace-preserving, Hermitian, non-CP map.

A linear map is called \textquotedblleft completely
positive\textquotedblright\ (CP) if it is a Hermitian map with $c_{\alpha
}\geq 0$ $\forall \alpha $. CP maps play a key role in quantum information
and quantum error correction \cite{Nielsen:book}, though they have a much
earlier origin \cite{Stinespring:55,Kraus:83}. There are other useful
characterizations of CP maps -- see, e.g., Refs.~\cite%
{Breuer:Book,Nielsen:book}. It turns out that there is a tight connection
between CP and Hermitian maps \cite{Jordan:04,Carteret:05}: a map is
Hermitian iff it can be written as the difference of two CP maps.

The definition of a CP map $\Phi _{\mathrm{CP}}$ implies that it can be
expressed in the Kraus operator sum representation \cite{Kraus:83}: 
\begin{equation}
\rho _{S}(t)=\sum_{\alpha }E_{\alpha }(t)\rho _{S}(0)E_{\alpha }^{\dagger
}(t)=\Phi _{\mathrm{CP}}(t)[\rho _{S}(0)].
\end{equation}%
If the operation elements $E_{\alpha }$ satisfy ${\sum_{\alpha }}E_{\alpha
}^{\dagger }E_{\alpha }=I$ then $\mathrm{Tr}[\rho _{S}(t)]=1$.

\subsection{Special linear states}

Following Ref. \cite{ShabaniLidar:08}, we define the class of
\textquotedblleft special-linear\textquotedblright\ (SL) states for which
the QDP (\ref{dynamics1}) always results in a linear, Hermitian map. An
arbitrary bipartite state on $\mathcal{H}_{S}\otimes \mathcal{H}_{B}$ can be
written as%
\begin{equation}
\rho _{SB}=\sum\nolimits_{ij}\varrho _{ij}|i\rangle \langle j|\otimes \phi
_{ij},  \label{eq:rho_SB}
\end{equation}%
where $\{|i\rangle \}_{i=1}^{\dim \mathcal{H}_{S}}$ is an orthonormal basis
for $\mathcal{H}_{S}$, and $\{\phi _{ij}\}_{i,j=1}^{\dim \mathcal{H}_{S}}:%
\mathcal{H}_{B}\mapsto \mathcal{H}_{B}$ are normalized such that if $\mathrm{%
Tr}[\phi _{ij}]\neq 0$ then $\mathrm{Tr}[\phi _{ij}]=1$. The corresponding
reduced system and bath states are then $\rho _{S}=\sum_{(i,j)\in \mathcal{C}%
}\varrho _{ij}|i\rangle \langle j|$, where $\mathcal{C}\equiv \{(i,j)|%
\mathrm{Tr}[\phi _{ij}]=1\}$, and $\rho _{B}(0)=\sum_{i}\varrho _{ii}\phi
_{ii}$. Hermiticity and normalization of $\rho _{SB}$, $\rho _{S}$, and $%
\rho _{B}$ imply $\varrho _{ij}=\varrho _{ji}^{\ast }$, $\phi _{ij}=\phi
_{ji}^{\dag }$, and $\sum_{i}\varrho _{ii}=1$.

\begin{mydefinition}
\label{def:SL}A bipartite state $\rho _{SB}$, parametrized as in Eq. (\ref%
{eq:rho_SB}), is in the SL-class\ if either $\mathrm{Tr}[\phi _{ij}]=1$ or $%
\phi _{ij}=0,$ $\forall i,j$.
\end{mydefinition}

Thus a non-SL\ state is a state for which there exist indexes $i$ and $j$
such that $\mathrm{Tr}[\phi _{ij}]=0$ but $\phi _{ij}\neq 0$. The following
result proven in Ref. \cite{ShabaniLidar:08} (generalizing an earlier result
in Ref. \cite{Rodriguez:07}) provides an almost complete answer to the
question posed above:

\begin{mytheorem}[Theorem 2 of \protect\cite{ShabaniLidar:08}]
\label{H-States}If $\rho _{SB}(0)$ is an SL-class state then the QDP (\ref%
{dynamics1}) is a linear, Hermitian map $\Phi _{\mathrm{H}}:\rho
_{S}(0)\mapsto \rho _{S}(t)$.
\end{mytheorem}

A further result proven in Ref. \cite{ShabaniLidar:08} (Theorem 3 there)
provides necessary and sufficient conditions on $\rho _{SB}(0)$ for the QDP (%
\ref{dynamics1}) to be a CP\ map, namely, $\rho _{SB}(0)$ should be a state
with vanishing quantum discord \cite{Ollivier:01}. Such a state cannot
contain any quantum correlations. This clearly illustrates the limitations
of CP\ maps in describing quantum dynamics. At the same time one may wonder
as to the generality of the SL-class employed in Theorem \ref{H-States}.
Non-SL\ states are sparse \cite{ShabaniLidar:08}, so it is in this regard
that we stated that Theorem \ref{H-States} provides an almost complete
answer to the question posed above. However, we can go further. As mentioned
without proof in Ref. \cite{ShabaniLidar:08}, in fact the QDP\ (\ref%
{dynamics1}) is a linear, Hermitian map from $\rho _{S}(0)\mapsto \rho
_{S}(t)$ for \emph{any} initial state $\rho _{SB}(0)$. We next prove this
key fact.

\subsection{Hermitian maps for arbitrary initial states}

We split the general initial state representation (\ref{eq:rho_SB}) into a
sum over SL\ and non-SL\ terms (thus splitting $\{\varrho _{ij}\}$ and $%
\{\phi _{ij}\}$ into two sets):%
\begin{equation}
\rho _{SB}(0)=\sum_{ij\in (\mathrm{SL})}\alpha _{ij}|i\rangle \langle
j|\otimes \varphi _{ij}+\sum_{ij\in (\mathrm{nSL})}\beta _{ij}|i\rangle
\langle j|\otimes \psi _{ij}.  \label{eq:SL}
\end{equation}%
In accordance with the definition of SL\ states, in the first sum we include
only terms $\alpha _{ij}|i\rangle \langle j|\otimes \varphi _{ij}$ for which 
$\mathrm{Tr}[\varphi _{ij}]\neq 0$ or $\varphi _{ij}=0$, in the second only
terms $\beta _{ij}|i\rangle \langle j|\otimes \psi _{ij}$ with bath
operators $\{\psi _{ij}\}$ satisfying $\psi _{ij}\neq 0$ and $\mathrm{Tr}%
[\psi _{ij}]=0$. By virtue of this decomposition only the first term
contributes to the initial system state: $\rho _{S}(0)=\mathrm{Tr}_{B}[\rho
_{SB}(0)]=\sum_{ij(\mathrm{SL})}\alpha _{ij}|i\rangle \langle j|$. This is
because the condition $\mathrm{Tr}[\psi _{ij}]=0$ eliminates any
contribution from the second term in the decomposition (\ref{eq:SL}) to the
initial system state. Consequently Eq. (\ref{eq:Qmap}) assumes an affine
form:%
\begin{equation}
\Phi _{\mathrm{Q}}(t)[\rho _{S}(0)]=\Phi _{\mathrm{SL}}(t)[\rho _{S}(0)]+K_{%
\mathrm{nSL}}(t),  \label{eq:affine}
\end{equation}%
with the term $K_{\mathrm{nSL}}(t)$ being a shift that is \emph{independent
of }$\rho _{S}(0)$.

As shown in Ref. \cite{ShabaniLidar:08}, the linear map $\Phi _{\mathrm{SL}}$
is constructed as a function of the bath operators $\{\varphi _{ij}\}$: 
\begin{equation}
\Phi _{\mathrm{SL}}(t)[\rho _{S}(0)]\equiv \sum_{(i,j)\in (\mathrm{SL}%
);k,\alpha }\lambda _{\alpha }^{ij}V_{kij}^{\alpha }P_{i}\rho
_{S}(0)P_{j}(W_{kij}^{\alpha })^{\dag },  \label{Linear}
\end{equation}%
where $P_{i}\equiv |i\rangle \langle i|$ are projectors, $\lambda _{\alpha
}^{ij}$ are the singular values in the singular value decomposition $\phi
_{ij}=\sum_{\alpha }\lambda _{\alpha }^{ij}|x_{ij}^{\alpha }\rangle \langle
y_{ij}^{\alpha }|$, and the operators $V_{kij}^{\alpha }\equiv \langle \psi
_{k}|U|x_{ij}^{\alpha }\rangle $ and $W_{kij}^{\alpha }\equiv \langle \psi
_{k}|U|y_{ij}^{\alpha }\rangle $ act on the system only, with $\{|\psi
_{k}\rangle \}$ being an orthonormal basis for the bath Hilbert space ${%
\mathcal{H}}_{B}$.

In addition, the non-SL terms in Eq.(\ref{eq:SL}) generate the shift term 
\begin{equation}
K_{\mathrm{nSL}}(t)=\sum_{ij\in (\mathrm{nSL})}\beta _{ij}\mathrm{Tr}%
_{B}[U_{SB}(t)|i\rangle \langle j|\otimes \psi _{ij}U_{SB}^{\dag }(t)].
\end{equation}%
This shows explicitly that $K_{\mathrm{nSL}}(t)$ does not depend on the
initial system state, since the latter is fully parametrized by the
coefficients $\{\alpha _{ij}\}_{ij\in (\mathrm{SL})}$, while $K_{\mathrm{nSL}%
}(t)$ depends only upon the coefficients $\{\beta _{ij}\}_{ij\in (\mathrm{nSL%
})}$.

Now we take a further step to argue that the affine map (\ref{eq:affine}) is
actually a linear, Hermitian map if the map acts only on the space of
density matrices. This is a direct application of the result in Ref. \cite%
{Jordan:04}.

\begin{mytheorem}
\label{linear Rep}The QDP (\ref{dynamics1}) is representable as a linear,
Hermitian map $\Phi _{\mathrm{H}}(t):\rho _{S}(0)\mapsto \rho _{S}(t)$ for
any initial system-bath state.
\end{mytheorem}

\begin{proof}
Let $N\equiv \dim \mathcal{H}_{S}$. Let $F_{0}\equiv I$ and let $\{F_{\mu }:%
\mathrm{Tr}(F_{\mu })=0\}_{\mu =1}^{N^{2}-1}$ be a basis for the set of
traceless Hermitian matrices which are mutually orthogonal with respect to
the Hilbert-Schmidt inner product, i.e., $\mathrm{Tr}(F_{\mu }F_{\nu
})=N\delta _{\mu \upsilon }$. Hence the initial system state $\rho _{S}(0)$
can be expanded as 
\begin{equation}
\rho _{S}(0)=\frac{1}{N}(I+\sum_{\mu =1}^{N^{2}-1}b_{\mu }F_{\mu });\quad
b_{\mu }=\mathrm{Tr}[\rho _{S}(0)F_{\mu }]\equiv \langle F_{\mu }\rangle
_{\rho _{S}(0)},  \label{vector}
\end{equation}%
and the final system state is found to be%
\begin{eqnarray}
\rho _{S}(t) &=&\frac{1}{N}[\Phi _{\mathrm{SL}}(I)+\sum_{\mu
=1}^{N^{2}-1}b_{\mu }\Phi _{\mathrm{SL}}(F_{\mu })]+K_{\mathrm{nSL}}  \notag
\\
&=&\Phi _{\mathrm{H}}(t)[\rho _{S}(0)],
\end{eqnarray}%
where the equivalent Hermitian map $\Phi _{\mathrm{H}}$ is constructed by
setting $\Phi _{\mathrm{H}}(I)=\Phi _{\mathrm{SL}}(I)+NK_{\mathrm{nSL}}$ and 
$\Phi _{\mathrm{H}}(F_{\mu })=\Phi _{\mathrm{SL}}(F_{\mu })$ $1\leq \forall
\mu \leq N^{2}-1$. That this map is Hermitian is simple to verify, for all
the components are Hermitian.
\end{proof}

Theorem \ref{linear Rep} provides a complete, and perhaps surprising answer
to the question posed at the beginning of this section. Namely, the most
general form of a quantum dynamical process, irrespective of the initial
system-bath state (in particular arbitrarily entangled initial states are
possible) is always reducible to a Hermitian map from the initial system to
the final system state. The surprising aspect of this result is that it was
not known previously whether QDP\ could always even be reduced to a \emph{map%
} between system states.

Of course, this result does not resolve the more difficult question of
ensuring the positivity of the final system state. That is, a Hermitian map
may transform an initially positive system state to a non-positive one,
violating the postulate of positivity of quantum states. To resolve this one
must identify the \textquotedblleft positivity domain\textquotedblright\ of $%
\Phi _{\mathrm{H}}$, i.e., the set of initial system states (positive by
definition) which are mapped to positive states by $\Phi _{\mathrm{H}}$ \cite%
{Jordan:04}. We address this in the next subsection.

\subsection{Geometric characterization of the Positivity Domain}

In this subsection we prove the convexity of the positivity domain and
propose a geometric method for characterizing it. Let $S(\mathcal{H})\equiv
\{\rho \in \mathcal{L}(\mathcal{H}):\rho >0,\mathrm{Tr}\rho =1\}$, where $%
\mathcal{L}(\mathcal{H})$ is the set of all linear operators on $\mathcal{H}$%
. The positivity domain of a linear map $\Phi _{\mathrm{L}}:S(\mathcal{H}%
)\mapsto \mathcal{B}(\mathcal{H})$ is: $P_{\Phi }\equiv \{\rho \in S(%
\mathcal{H}_{S}):\Phi _{\mathrm{L}}(\rho )>0\}$. 

Following earlier work \cite{Jakobczyk:01,Kimura:03,byrd:062322}, in Ref.~%
\cite{Kimura:05}, a complete geometric characterization of density matrices
was given by using the Bloch vector representation for an arbitrary $N$%
-dimensional Hilbert space $\mathcal{H}$. This works as follows:\ let $%
\{F_{\mu }\}_{\mu =1}^{N^{2}-1}$ be a basis set as in the proof of Theorem %
\ref{linear Rep}, whence the expansion (\ref{vector}) applies again. The
vector $\mathbf{b}=(b_{1},...,b_{N^{2}-1})\in \mathbb{R}^{N^{2}-1}$ of
expectation values is known as the Bloch vector, and knowing its components
is equivalent to complete knowledge of the corresponding density matrix, via
the map $\mathbf{b}\mapsto \rho \mathbf{=}\frac{1}{N}(I+\sum_{\mu
=1}^{N^{2}-1}b_{\mu }F_{\mu })$. Let $\mathbf{n}$ denote a unit vector,
i.e., $\mathbf{n}\in \mathbb{R}^{N^{2}-1}$ and $%
\sum_{i=1}^{N^{2}-1}n_{i}^{2}=1$, and define $F_{\mathbf{n}}\equiv \sum_{\mu
=1}^{N^{2}-1}n_{\mu }F_{\mu }$. Let the minimum eigenvalue of each $F_{%
\mathbf{n}}$ be denoted $m(F_{\mathbf{n}})$. The \textquotedblleft Bloch
space\textquotedblright\ $\boldsymbol{B}(\mathbb{R}^{N^{2}-1})$ is the set
of all Bloch vectors and is a closed convex set, since the set $S(\mathcal{H}%
)$ is closed and convex, and the map $\mathbf{b}\mapsto \rho $ is linear
homeomorphic. As shown in Theorem 1 of Ref.~\cite{Kimura:05}, the Bloch
space is characterized in the \textquotedblleft spherical
coordinates\textquotedblright\ determined by $\{F_{\mathbf{n}}\}$ as:%
\begin{equation}
\boldsymbol{B}(\mathbb{R}^{N^{2}-1})=\left\{ \mathbf{b}=r\mathbf{n}\in 
\mathbb{R}^{N^{2}-1}:r\leq \frac{1}{|m(F_{\mathbf{n}})|}\right\} .
\label{eq:Bloch-space}
\end{equation}%
It is hard to imagine a more intuitive or simpler geometric picture. 

Next we show that the positivity domain is a convex set as well.

\begin{myproposition}
The positivity domain $P_{\Phi }$ of a linear map $\Phi _{\mathrm{L}}$ is a
convex set.
\end{myproposition}

\begin{proof}
Consider two density matrices $\rho $ and $\rho ^{\prime }$ as interior
points of $P_{\Phi }$ with corresponding Bloch vectors $\mathbf{b}%
=(b_{1},...,b_{N^{2}-1})$ and $\mathbf{b^{\prime }}=(b_{1}^{\prime
},...,b_{N^{2}-1}^{\prime })$. The claim is that a third density matrix $%
\rho ^{\prime \prime }$ with corresponding Bloch vector $\mathbf{b^{\prime
\prime }}(\alpha )=\alpha \mathbf{b}+(1-\alpha )\mathbf{b^{\prime }}$, with $%
0\leq \alpha \leq 1$, is then also interior to $P_{\Phi }$. This follows
directly by linearity of the map $\Phi _{\mathrm{L}}$. First, by assumption $%
\Phi _{\mathrm{L}}[\rho ]=\Phi _{\mathrm{L}}[\frac{1}{N}(I+\sum_{\mu
=1}^{N^{2}-1}b_{\mu }F_{\mu })]>0$ and $\Phi _{\mathrm{L}}[\rho ^{\prime
}]=\Phi _{\mathrm{L}}[\frac{1}{N}(I+\sum_{\mu =1}^{N^{2}-1}b_{\mu }^{\prime
}F_{\mu })]>0$, so that $\alpha \Phi _{\mathrm{L}}[\rho ]+(1-\alpha )\Phi _{%
\mathrm{L}}[\rho ^{\prime }]>0$. Second, $\alpha \Phi _{\mathrm{L}}[\rho
]+(1-\alpha )\Phi _{\mathrm{L}}[\rho ^{\prime }]=\Phi _{\mathrm{L}}[\frac{1}{%
N}I]+\alpha \sum_{\mu =1}^{N^{2}-1}b_{\mu }\Phi _{\mathrm{L}}[F_{\mu
}]+(1-\alpha )\sum_{\mu =1}^{N^{2}-1}b_{\mu }^{\prime }\Phi _{\mathrm{L}%
}[F_{\mu }]=\Phi _{\mathrm{L}}[\frac{1}{N}(I+\sum_{\mu =1}^{N^{2}-1}b_{\mu
}^{\prime \prime }F_{\mu })]=\Phi _{\mathrm{L}}[\rho ^{\prime \prime }]$.
Therefore indeed $\Phi _{\mathrm{L}}[\rho ^{\prime \prime }]>0$.
\end{proof}

We are now ready to describe an algorithm for finding the boundary of the
positivity domain $P_{\Phi }$. We know at this point that $P_{\Phi }$ is
convex and that $P_{\Phi }$ is a subset of the Bloch space, itself a closed
convex set. Pick a unit vector $\mathbf{n}$ and draw a line through the
origin of the Bloch space along $\mathbf{n}$.
If $P_{\Phi }$
includes the origin, i.e., the maximally mixed state, then convexity implies that
this line intersects the boundary of $P_{\Phi }$ once. If $P_{\Phi }$
does not include the origin then convexity implies that this line
either intersects the boundary of $P_{\Phi }$ twice or not at all.
I.e., it follows from convexity that
the line may not re-enter the positivity domain once it exited. In order to
determine this boundary we may thus compute the eigenvalues of $\Phi _{%
\mathrm{L}}[\rho _{\mathbf{n}}(r)]$ as a function of $r$, where $r$ is the
parameter in Eq. (\ref{eq:Bloch-space}), and where $\rho _{\mathbf{n}}(r)$
is the density matrix determined via the mapping $\mathbf{b}=r\mathbf{n}%
\mapsto \rho $. The computation should start from $r=0$ and go up to at most 
$r=1/|m(F_{\mathbf{n}})|$. The boundary is identified as soon as the
eigenvalues of $\Phi _{\mathrm{L}}[\rho _{\mathbf{n}}(r)]$ go from all positive semi-definite to at least one negative, or vice versa. For
each unit vector $\mathbf{n}$, the corresponding point on the border of the
positivity domain can be found in this way. Then the algorithm constructs
the boundary of the positivity domain by finding the boundary points in all
directions $\mathbf{n}$. Of course, in practice one can only sample the
space of unit vectors $\mathbf{n}$ and factors $r$. In principle this yields
a complete geometrical description of the positivity domain of a given
linear map.

\section{CP\ maps and fault tolerant quantum error correction}

\label{sec:FTQEC}

\subsection{CP\ maps: pro and con}

We have already mentioned that a QDP (\ref{dynamics1}) becomes a CP map iff
the initial system-bath state has vanishing quantum discord, i.e., is purely
classically correlated \cite{ShabaniLidar:08}. The standard argument in
favor of CP\ maps is that since the system $S$ may be coupled with the bath $%
B$, the maps describing physical processes on $S$ should be such that all
their extensions into higher dimensional spaces should remain positive,
i.e., $\Phi _{\mathrm{CP}}\otimes I_{n}\geq 0$ $\forall n\in \mathbb{Z}^{+}$%
, where $I_{n}$ is the $n$-dimensional identity operator. However, one may
question whether this is the right criterion for describing quantum dynamics 
\cite{Pechukas+Alicki:95}. An alternative viewpoint is to seek a description
that applies to \emph{arbitrary} $\rho _{SB}(0)$, as we have done above. We
now argue that this viewpoint is the correct one for fault-tolerant quantum
error correction (FT-QEC).

\subsection{(In)validity of the CP\ map model in FT-QEC}

Let us show that system-environment correlations impose a severe restriction
on the applicability of CP maps in FT-QEC. The CP map model used in FT-QEC 
\cite%
{Shor:96,Aharonov:96,Knill:98,Steane:03,Knill:05,Reichardt:05,Aharonov:08,Aliferis:08}
can be described as follows (see, e.g., Eq. (8.1) in \cite{Aharonov:08}): $%
\rho _{S}(T)=\Phi _{\mathrm{CP}}^{\mathrm{tot}}(T,t_{0})[\rho _{S}(t_{0})]$
where 
\begin{equation}
\Phi _{\mathrm{CP}}^{\mathrm{tot}}(T,t_{0})=\bigotimes\nolimits_{i=1}^{N}%
\Phi _{U}(t_{i})\Phi _{\mathrm{CP}}(t_{i},t_{i-1}),  \label{eq:mapmodel}
\end{equation}%
where $T\equiv t_{N}$ is the total circuit time, and where $\Phi _{U}[\rho
_{S}]=U_{S}\rho _{S}U_{S}^{\dagger }$ is a unitary map (automatically CP)
that describes an ideal quantum logic gate.\footnote{%
In this subsection we denote noise maps by their initial and final times, to
distinguish them from the instantaneous unitary maps.} This represents the
idea used repeatedly in FT-QEC, that the noisy evolution at every time step
can be decomposed into \textquotedblleft pure noise\textquotedblright\ $\Phi
_{\mathrm{CP}}(t_{i},t_{i-1})$ followed by an instantaneous and perfect
unitary gate $\Phi _{U}(t_{i})$. More precisely, in FT-QEC one assumes that
the evolution starts ($t=t_{0}=0$) from a product state, then undergoes a
CP\ map $\Phi _{\mathrm{CP}}(t_{1},t_{0})$ due to coupling to the
environment, followed by an instantaneous error correction step $\Phi
_{U}(t_{1})$. If the latter were perfect then the post-error-correction
state would again be a product state $\rho _{S}(t_{1})\otimes \rho
_{B}(t_{1})$. However, FT-QEC\ allows for the fact that the error correction
step is almost never perfect, which means that there is a residual
correlation between system and bath at $t_{1}$. Hence, according to Ref.~%
\cite{ShabaniLidar:08}, the map that describes the evolution of the system
is a CP\ map if and only if the residual correlation is purely classical.
Otherwise it is a Hermitian map. To make this point more explicit, consider
a sequence of two noise time-steps, interrupted by one error correction
step. In the ideal scenario, where the error correction step $\Phi
_{U}(t_{1})$ works perfectly (i.e., reduces the system-bath correlations to
purely classical), we would have%
\begin{equation}
\Phi _{\mathrm{CP}}^{\mathrm{(2)}}(t_{2},t_{0})=\Phi _{\mathrm{CP}%
}(t_{2},t_{1})\Phi _{U}(t_{1})\Phi _{\mathrm{CP}}(t_{1},t_{0}),
\end{equation}%
where $\Phi _{\mathrm{CP}}(t_{2},t_{1})$ is again a CP\ noise map. However,
in reality $\Phi _{U}(t_{1})$ works imperfectly [system-bath correlations
are not purely classical after the action of $\Phi _{U}(t_{1})$], and the
actual map obtained is%
\begin{equation}
\Phi _{\mathrm{H}}^{\mathrm{(2)}}(t_{2},t_{0})=\Phi _{\mathrm{H}%
}(t_{2},t_{1})\Phi _{U}(t_{1})\Phi _{\mathrm{CP}}(t_{1},t_{0}),
\end{equation}%
where $\Phi _{\mathrm{H}}(t_{2},t_{1})$ is now a Hermitian map. Note that,
in fact, even the assumption that the first noise map is CP will not be true
in general, due to errors in the preparation of the initial state, leading
to non-classical correlations between system and bath. We conclude that in
general the CP\ map model (\ref{eq:mapmodel}) should be replaced by 
\begin{equation}
\Phi _{\mathrm{H}}^{\mathrm{tot}}(T,t_{0})=\bigotimes\nolimits_{i=1}^{N}\Phi
_{U}(t_{i})\Phi _{\mathrm{H}}(t_{i},t_{i-1}),  \label{eq:linmap}
\end{equation}%
where $\Phi _{\mathrm{H}}(t_{i},t_{i-1})$ are \emph{Hermitian maps}, not
necessarily CP.\footnote{%
Note that Eq.~(\ref{eq:linmap}) applies also to non-Markovian noise, and is
hence complementary to Hamiltonian FT-QEC \cite%
{Terhal:04,Aliferis:05,Aharonov:05}.}

It is worth emphasizing that this distinction between purely classical and
other correlations, and the resulting difference between CP\ and Hermitian
evolution, is not a distinction that has thus far been made in FT-QEC\
theory. Rather, in FT-QEC one distinguishes between \textquotedblleft
good\textquotedblright\ and \textquotedblleft bad\textquotedblright\ fault
paths, where the former (latter) contain only a few (too many) errors.
Quoting from \cite{Terhal:04}: \textquotedblleft There are good fault paths
with so-called sparse numbers of faults which keep being corrected during
the computation and which lead to (approximately) correct answers of the
computation; and there are bad fault-paths which contain too many faults to
be corrected and imply a crash of the quantum computer.\textquotedblright\
This leads to a splitting of the total map (\ref{eq:mapmodel}) into a sum
over good and bad paths. One then shows that the computation can proceed
robustly via the use of concatenated codes, provided the \textquotedblleft
bad\textquotedblright\ paths are appropriately bounded. In \cite{Aharonov:08}%
(p.1272) it was pointed out that the sum over \textquotedblleft
good\textquotedblright\ paths need not be a CP\ map, but can be decomposed
into a new sum over CP\ maps [Eq. (8.13) there]. This new decomposition can
then be treated using standard FT-QEC techniques. However, this assumes
again that the total evolution is a CP\ map, which in fact it is not [Eq. (%
\ref{eq:linmap})].

These observations motivate a generalized theory of QEC, which can handle
non-CP\ noise maps. This is the subject of the next section. \emph{The main
result of this theory is reassuring:\ in spite of the invalidity of the CP\
map model in FT-QEC, the CP-map based results apply because the same
encoding and recovery that corrects a Hermitian map can be used to correct a
closely related CP\ map, whose coefficients are the absolute values of the
Hermitian map}. This is formalized in Corollary \ref{cor:HM}.

\section{ Linear Quantum Error Correction}

\label{sec:LQEC}

Having argued that non-CP Hermitian maps arise naturally in the study of
open systems, and in particular FT-QEC, we now proceed to develop the theory
of Linear QEC. For generality we do this for arbitrary linear maps, i.e.,
maps of the form (\ref{eq:LM}). We then specialize to the physically
relevant case of Hermitian maps.

Let us first recall the fundamental theorem of \textquotedblleft
standard\textquotedblright\ QEC (for CP noise and CP recovery maps) \cite%
{Knill:97b}: Let $P$ be a projection operator onto the code space. Necessary
and sufficient conditions for quantum error correction of a CP map, $\Phi _{%
\mathrm{CP}}(\rho )={\sum_{i}}F_{i}\rho F_{i}^{\dagger }$ are 
\begin{equation}
PF_{i}^{\dag }F_{j}P=\lambda _{ij}P\quad \forall i,j.  \label{eq:QEC-CP}
\end{equation}%
An elegant proof of this theorem and a construction of the corresponding CP
recovery map was given in Refs. \cite{Nielsen:98,Nielsen:book}; we use some
of their methods in the proofs of Theorems~\ref{th:CP-rec},\ref{th:suff}.

\subsection{CP-recoverable linear noise maps}

While general (non-Hermitian) linear maps of the form (\ref{eq:LM}) do not
arise from quantum dynamical processes [Eq. (\ref{dynamics1})], it is still
interesting from a purely mathematical standpoint to consider QEC for such
maps. Moreover, we easily recover the physical setting from these general
considerations.

Theorem~\ref{th:CP-rec} shows that there is a class of linear noise maps
which are equivalent to certain non-trace-preserving CP noise maps when it
comes to error correction using CP recovery maps.

\begin{mytheorem}
\label{th:CP-rec}Consider a general linear noise map $\Phi _{\mathrm{L}%
}(\rho )$ $={\sum_{i=1}^{N}}E_{i}\rho E_{i}^{\prime \dagger }$ and associate
to it an \textquotedblleft expanded\textquotedblright\ CP map $\tilde{\Phi}_{%
\mathrm{CP}}(\rho )=\frac{1}{2}{\sum_{i=1}^{N}}E_{i}\rho E_{i}^{\dagger }+%
\frac{1}{2}{\sum_{i=1}^{N}}E_{i}^{\prime }\rho E_{i}^{\prime \dagger }$.
Then any QEC code $\mathcal{C}$ and corresponding CP recovery map $\mathcal{R%
}$ for $\tilde{\Phi}_{\mathrm{CP}}$ are also a QEC code and CP recovery map
for $\Phi _{\mathrm{L}}$.
\end{mytheorem}

\begin{proof}
The operation elements of $\tilde{\Phi}_{\mathrm{CP}}$ are $%
\{F_{i}\}_{i=1}^{N}=\{\frac{1}{\sqrt{2}}E_{i}\}_{i=1}^{N}$ and $%
\{F_{N+i}\}_{i=1}^{N}=\{\frac{1}{\sqrt{2}}E_{i}^{\prime }\}_{i=1}^{N}$,
whence $\tilde{\Phi}_{\mathrm{CP}}(\rho )={\sum_{i=1}^{2N}}F_{i}\rho
F_{i}^{\dagger }$. The standard quantum error correction conditions (\ref%
{eq:QEC-CP}) for $\tilde{\Phi}_{\mathrm{CP}}$, where 
\begin{equation}
\lambda \equiv 2\left( 
\begin{array}{cc}
\alpha & \gamma \\ 
\gamma ^{\dag } & \alpha ^{\prime }%
\end{array}%
\right) =\lambda ^{\dag },  \label{eq:lambda}
\end{equation}%
become three sets of conditions in terms of the $E_{i}$ and $E_{i}^{\prime }$%
: 
\begin{eqnarray}
\text{(i)}~PE_{i}^{\dag }E_{j}P &=&2\alpha _{ij}P,~\text{(ii)}%
~PE_{i}^{\prime \dag }E_{j}^{\prime }P=2\alpha _{ij}^{\prime }P,  \notag \\
\text{(iii)}~PE_{i}^{\dag }E_{j}^{\prime }P &=&2\gamma _{ij}P,
\label{eq:lambda(i)}
\end{eqnarray}%
where $i,j\in \{1,...,N\}$ and $\alpha _{ij}=\lambda _{ij}$, $\gamma
_{ij}=\lambda _{i,N+j}$, $\alpha _{ij}^{^{\prime }}=\lambda _{N+i,N+j}$. The
existence of a projector $P$ which satisfies Eqs.~(\ref{eq:lambda(i)}%
)(i)-(iii) is equivalent to the existence of a QEC code for $\tilde{\Phi}_{%
\mathrm{CP}}$. Assuming that a code $\mathcal{C}$ has been found (i.e., $P%
\mathcal{C}=\mathcal{C}$) for $\tilde{\Phi}_{\mathrm{CP}}$, we use this as a
code for $\Phi _{\mathrm{L}}$ and show that the corresponding CP recovery
map $\mathcal{R}_{\mathrm{CP}}$ is also a recovery map for $\Phi _{\mathrm{L}%
}$. Indeed, let $G_{j}\equiv \sum_{i=1}^{2N}u_{ij}F_{i}$ be new operation
elements for $\tilde{\Phi}_{\mathrm{CP}}$, where $u$ is the unitary matrix
that diagonalizes $\lambda $, i.e., $u^{\dag }\lambda u=d$. Then $\tilde{\Phi%
}_{\mathrm{CP}}={\sum_{j=1}^{2N}}G_{j}\rho G_{j}^{\dagger }$. Let $\mathcal{R%
}_{\mathrm{CP}}=\{R_{k}\}$ be the CP recovery map for $\tilde{\Phi}_{\mathrm{%
CP}}$. Assume that $\rho $ is in the code space, i.e., $P\rho P=\rho $. We
now show that $\mathcal{R}_{\mathrm{CP}}[\Phi _{\mathrm{L}}(\rho )]=\rho $,
i.e., we have CP recovery. First, 
\begin{eqnarray}
\mathcal{R}_{\mathrm{CP}}[\Phi _{\mathrm{L}}(\rho )] &=&\sum_{k}R_{k}\left(
\sum_{i=1}^{N}F_{i}\rho F_{N+i}^{\dag }\right) R_{k}^{\dag }  \notag \\
&=&\sum_{i=1}^{N}\sum_{j,j^{\prime }=1}^{2N}u_{ij}^{\ast }u_{N+i,j^{\prime
}}\times  \notag \\
&&\sum_{k}\left( R_{k}G_{j}P\right) \rho \left( PG_{j^{\prime }}^{\dag
}R_{k}^{\dag }\right) .  \label{eq:R-CP0}
\end{eqnarray}%
Now, note that 
\begin{eqnarray}
PG_{k}^{\dag }G_{l}P &=&\sum_{ij}u_{ik}^{\ast }u_{jl}PF_{i}^{\dag
}F_{j}P=\sum_{ij}u_{ik}^{\ast }\lambda _{ij}u_{jl}P  \notag \\
&=&d_{k}\delta _{kl}P.
\end{eqnarray}%
Then the polar decomposition yields 
\begin{equation}
G_{k}P=U_{k}(PG_{k}^{\dag }G_{k}P)^{1/2}=\sqrt{d_{k}}U_{k}P.
\end{equation}%
The recovery operation elements are given by 
\begin{equation}
R_{k}=U_{k}^{\dag }P_{k},  \label{eq:recovery}
\end{equation}%
where $P_{k}=U_{k}PU_{k}^{\dag }$. Therefore $P_{k}=G_{k}PU_{k}^{\dag }/%
\sqrt{d_{k}}$. This allows us to calculate the action of the $k$th recovery
operator on the $l$th error \cite{Nielsen:98,Nielsen:book}: 
\begin{eqnarray}
R_{k}G_{l}P &=&U_{k}^{\dag }P_{k}^{\dag }G_{l}P=U_{k}^{\dag
}(U_{k}PG_{k}^{\dag }/\sqrt{d_{k}})G_{l}P  \notag \\
&=&\delta _{kl}\sqrt{d_{k}}P.
\end{eqnarray}%
Therefore, 
\begin{eqnarray}
\mathcal{R}_{\mathrm{CP}}[\Phi _{\mathrm{L}}(\rho )]
&=&\sum_{i=1}^{N}\sum_{j,j^{\prime }=1}^{2N}u_{ij}^{\ast }u_{N+i,j^{\prime }}
\notag \\
&&\times \sum_{k}\left( \delta _{kj}\sqrt{d_{k}}P\right) \rho \left( P\sqrt{%
d_{k}}\delta _{kj^{\prime }}\right)  \notag \\
&=&\rho \sum_{i=1}^{N}\left( udu^{\dag }\right) _{N+i,i}=\rho
\sum_{i=1}^{N}\lambda _{N+i,i}  \notag \\
&=&2\rho \mathrm{Tr}\gamma ^{\dag }.  \label{eq:R-CP}
\end{eqnarray}%
Next note that, using condition (\ref{eq:lambda(i)})(iii) and trace
preservation by $\Phi _{\mathrm{L}}$: 
\begin{eqnarray}
PE_{i}^{\prime \dag }E_{i}P &=&2\gamma _{ii}^{\dag }P\Longrightarrow 2%
\mathrm{Tr}\gamma ^{\dag }P=P\sum_{i}E_{i}^{\prime \dag }E_{i}P=P  \notag \\
&\Longrightarrow &\mathrm{Tr}\gamma ^{\dag }=\frac{1}{2}.
\end{eqnarray}%
Hence, finally: 
\begin{equation}
\mathcal{R}_{\mathrm{CP}}[\Phi _{\mathrm{L}}(\rho )]=\rho  \label{eq:RL_CP}
\end{equation}%
for any $\rho $ in the codespace.
\end{proof}

Note that $\tilde{\Phi}_{\mathrm{CP}}(\rho )$ need not be trace preserving:$%
\ \mathrm{Tr}[\tilde{\Phi}_{\mathrm{CP}}(\rho )]=\frac{1}{2}\mathrm{Tr}[({%
\sum_{i=1}^{N}}E_{i}^{\dagger }E_{i}+{\sum_{i=1}^{N}}E_{i}^{\prime \dagger
}E_{i}^{\prime })\rho ]$, and while ${\sum_{i=1}^{N}}E_{i}^{\prime \dagger
}E_{i}=I$ if $\Phi _{\mathrm{L}}$ is trace preserving, we do not have
conditions on ${\sum_{i=1}^{N}}E_{i}^{\dagger }E_{i}$ and ${\sum_{i=1}^{N}}%
E_{i}^{\prime \dagger }E_{i}^{\prime }$.

We define the class of \textquotedblleft CP-recoverable linear noise
maps\textquotedblright\ $\{\Phi _{\mathrm{CPR}}\}$ as those $\Phi _{\mathrm{L%
}}$ for which CP\ recovery is always possible. By Theorem~\ref{th:CP-rec}
this includes all $\Phi _{\mathrm{L}}$ for which $P$ can be found satisfying
conditions~(\ref{eq:lambda(i)})(i)-(iii). However, these conditions are not
necessary.

\subsection{Non-CP-recoverable linear noise maps}

We now define \textquotedblleft non-CP-recoverable linear noise
maps\textquotedblright\ $\{\Phi _{\mathrm{nCPR}}\}$ as those $\Phi _{\mathrm{%
L}}$ for which non-CP-recovery is always possible. Theorem~\ref{th:suff}
shows constructively that $\{\Phi _{\mathrm{nCPR}}\}$ includes all linear
noise maps $\Phi _{\mathrm{L}}$ for which $P$ can be found satisfying only
conditions~(\ref{eq:lambda(i)})(i) and (ii). Clearly, $\{\Phi _{\mathrm{CP}%
}\}\subset \{\Phi _{\mathrm{CPR}}\}\subset \{\Phi _{\mathrm{nCPR}}\}\subset
\{\Phi _{\mathrm{L}}\}$.

\begin{mytheorem}
\label{th:suff} Let $\Phi _{\mathrm{L}}=\{E_{i},E_{i}^{\prime }\}_{i}$ be a
linear noise map. Then every state $\rho =P\rho P$ encoded using a QEC code
defined by a\ projector $P$\ satisfying only Eqs.~(\ref{eq:lambda(i)})(i)
and (ii) can be recovered using a non-CP recovery map.
\end{mytheorem}

\begin{proof}
Let $G_{k}=\sum_{i}u_{ik}E_{i}$ and $G_{k}^{\prime }=\sum_{i}u_{ik}^{\prime
}E_{i}^{\prime }$, where the unitaries $u$ and $u^{\prime }$ respectively
diagonalize the Hermitian matrices $\alpha $ and $\alpha ^{\prime }$: $%
d=u^{\dag }\alpha u$ and $d^{\prime }=u^{\prime \dag }\alpha ^{\prime
}u^{\prime }$. Define a recovery map $\mathcal{R}=\{R_{k},R_{k}^{\prime }\}$
(not necessarily CP) with operation elements%
\begin{equation}
R_{k}=U_{k}^{\dag }P_{k},\quad R_{k}^{\prime }=U_{k}^{\prime \dag
}P_{k}^{\prime }.  \label{eq:R}
\end{equation}%
Here $P_{k}=U_{k}PU_{k}^{\dag }$, $P_{k}^{\prime }=U_{k}^{\prime
}PU_{k}^{\prime \dag }$ are projection operators, and $U_{k}$ and $%
U_{k}^{\prime }$ arise from the polar decomposition of $G_{k}P$ and $%
G_{k}^{\prime }P$, i.e., $G_{k}P=U_{k}(PG_{k}^{\dag }G_{k}P)^{1/2}$ and $%
G_{k}^{\prime }P=U_{k}(PG_{k}^{\prime \dag }G_{k}^{\prime }P)^{1/2}$. The
proof is entirely analogous to the proof of Theorem~\ref{th:CP-rec}, except
that we must keep track of both the primed and unprimed operators. Following
through the same calculations we thus obtain $R_{k}G_{l}\sqrt{\rho }=\sqrt{%
d_{k}}\delta _{kl}\sqrt{\rho }$ and $R_{k}^{\prime }G_{l}^{\prime }\sqrt{%
\rho }=\sqrt{d_{k}^{\prime }}\delta _{kl}\sqrt{\rho }$. Using this in the
recovery map applied to the linear noise map, we find: 
\begin{eqnarray}
\mathcal{R}[\Phi (P\rho P)] &=&\sum_{kl}R_{k}E_{l}P\rho PE_{l}^{\prime
\dagger }R_{k}^{\prime \dag }  \notag \\
&=&\sum_{kl}R_{k}(\sum_{j}u_{lj}^{\ast }G_{j})P\rho P(\sum_{i}u_{li}^{\prime
}G_{i}^{\prime \dag })R_{k}^{\prime \dag }  \notag \\
&=&F_{\mathrm{L}}P\rho P\propto \rho ,
\end{eqnarray}%
where 
\begin{eqnarray}
F_{\mathrm{L}} &\equiv &\sum_{ijkl}u_{lj}^{\ast }u_{li}^{\prime }\sqrt{%
d_{k}d_{k}^{\prime \ast }}\delta _{kj}\delta _{ki}=\sum_{kl}u_{lk}^{\ast
}u_{lk}^{\prime }\sqrt{d_{k}d_{k}^{\prime \ast }}  \notag \\
&=&\mathrm{Tr}[u^{\prime }d^{\prime \dag }du^{\dag }]=\mathrm{Tr}[u^{\prime
}u^{\dag }\alpha \alpha ^{\prime \dag }]  \label{eq:F_L}
\end{eqnarray}%
is a \textquotedblleft correction factor\textquotedblright\ for non-CP\
recovery of linear noise maps, which was $1$ in the case of CP\ recovery,
above.
\end{proof}

Gathering the expressions derived in the last proof, we have the following
explicit expressions for the left and right recovery operations:%
\begin{equation}
R_{k}=U_{k}^{\dag }P_{k}^{\dag }=\frac{1}{\sqrt{d_{k}}}P\sum_{i}u_{ik}^{\ast
}E_{i}^{\dag },~R_{k}^{\prime }=\frac{1}{\sqrt{d_{k}^{\prime }}}%
P\sum_{i}u_{ik}^{\prime \ast }E_{i}^{\prime \dag }.
\end{equation}%
This also shows that, in general, $R_{k}$ need not equal $R_{k}^{\prime }$,
i.e., the recovery map is linear but not necessarily CP.

Note that standard QEC can also be interpreted as \textquotedblleft error
correction by inversion\textquotedblright , in the following sense: when the
noise map is CP and recovery is also CP, recovery is the inverse of the
noise map \emph{restricted to the code space} (Theorem III.3 in Ref.~\cite%
{Knill:97b}). The same is true for our LQEC results above, which relax the
restriction to CP noise maps.

\subsection{The physical case: Hermitian maps}

The general physical case is the case of Hermitian noise maps, to which any
quantum dynamical process can be reduced, as follows from Theorem \ref%
{linear Rep}. We can specialize Theorems \ref{th:CP-rec} and \ref{th:suff}
to this case.

\begin{mycorollary}
\label{cor:HM}Consider a Hermitian noise map $\Phi _{\mathrm{H}}(\rho )$ $={%
\sum_{i=1}^{N}}c_{i}K_{i}\rho K_{i}^{\dagger }$ and associate to it a CP map 
$\tilde{\Phi}_{\mathrm{CP}}(\rho )={\sum_{i=1}^{N}|}c_{i}|K_{i}\rho
K_{i}^{\dagger }$. Then any QEC code $\mathcal{C}$ and corresponding CP
recovery map $\mathcal{R}_{\mathrm{CP}}$ for $\tilde{\Phi}_{\mathrm{CP}}$
are also a QEC code and CP recovery map for $\Phi _{\mathrm{H}}$.
\end{mycorollary}

The important conclusion we can draw from Corollary \ref{cor:HM} is that
standard QEC\ techniques apply whether the noise map is CP\ or, as it will
almost always be due to non-classical correlations, Hermitian. This is
because Corollary \ref{cor:HM} tells us that it is safe to replace all
negative $c_{i}$ coefficients by their absolute values, and thus replace the
actual noise map by its CP\ counterpart.

\begin{proof}
We have $\Phi _{\mathrm{H}}(\rho )={\sum_{i=1}^{N}}E_{i}\rho E_{i}^{\prime
\dagger }$ with $\{{E}_{i}=\sqrt{c_{i}}K_{i}\}_{i=1}^{N}$ and $\{{E}%
_{i}^{\prime }=(\sqrt{c_{i}})^{\ast }K_{i}\}_{i=1}^{N}$, whence we can apply
the construction of Theorem \ref{th:CP-rec}. Indeed, the \textquotedblleft
expanded\textquotedblright\ CP map becomes $\tilde{\Phi}_{\mathrm{CP}}(\rho
)=\frac{1}{2}{\sum_{i=1}^{N}}E_{i}\rho E_{i}^{\dagger }+\frac{1}{2}{%
\sum_{i=1}^{N}}E_{i}^{\prime }\rho E_{i}^{\prime \dagger }={\sum_{i=1}^{N}|}%
c_{i}|K_{i}\rho K_{i}^{\dagger }$, as claimed, and hence a QEC\ code and CP\
recovery for $\tilde{\Phi}_{\mathrm{CP}}$ is also a QEC\ code and CP\
recovery for $\Phi _{\mathrm{H}}$. In particular, $\mathcal{R}_{\mathrm{CP}%
}[\Phi _{\mathrm{H}}(\rho )]=\rho $.
\end{proof}

Note that $\tilde{\Phi}_{\mathrm{CP}}$ need not be trace preserving even in
the Hermitian map case: $\mathrm{Tr}[\tilde{\Phi}_{\mathrm{CP}}(\rho )]=%
\mathrm{Tr}[{\sum_{i=1}^{N}|}c_{i}|K_{i}^{\dagger }K_{i}\rho ]$, but if $%
\Phi _{\mathrm{H}}$ is trace preserving then we only have ${\sum_{i=1}^{N}}%
c_{i}K_{i}^{\dagger }K_{i}=I$, hence cannot conclude more about $\mathrm{Tr}[%
\tilde{\Phi}_{\mathrm{CP}}(\rho )]$. Also note that substitution of ${E}_{i}=%
\sqrt{c_{i}}K_{i}$ and ${E}_{i}^{\prime }=(\sqrt{c_{i}})^{\ast }K_{i}$ into
the QEC conditions (\ref{eq:lambda(i)})(i)-(iii) yields $\alpha
_{ij}^{\prime }=\sqrt{\frac{c_{i}}{c_{j}}}\left( \sqrt{\frac{c_{j}}{c_{i}}}%
\right) ^{\ast }\alpha _{ij}$ and $\gamma _{ij}=\frac{(\sqrt{c_{j}})^{\ast }%
}{\sqrt{c_{j}}}\alpha _{ij}$, i.e., unlike in the general linear maps case,
the matrices $\alpha ^{\prime }$ and $\gamma $ in Eq. (\ref{eq:lambda}) are
not independent from $\alpha $. In fact, as shown in Appendix~\ref%
{app:direct} we can give a direct proof of Corollary \ref{cor:HM} which only
invokes a single block of the $\lambda $ matrix.

\subsubsection{Example of CP recovery: Inverse bit-flip map}

Consider \textquotedblleft diagonalizable maps\textquotedblright , i.e., $%
\Phi _{\mathrm{D}}(\rho )\equiv \sum_{i}c_{i}K_{i}\rho K_{i}^{\dagger }$,
where $c_{i}\in \mathbb{C}$. The expanded CP map is $\tilde{\Phi}_{\mathrm{CP%
}}=\sum_{i}|c_{i}|K_{i}\rho K_{i}^{\dagger }$. Now consider as a specific
instance an independent-errors inverse bit-flip map on three qubits: $\Phi _{%
\mathrm{IPF}}(\rho )=c_{0}\rho +c_{1}\sum_{n=1}^{3}X_{n}\rho X_{n}$, where $%
X_{n}$ is the Pauli $\sigma _{x} $ matrix applied to qubit $n$, where $c_{0}$
and $c_{1}$ are real, have opposite sign, and $c_{0}+3c_{1}=1$ (a Hermitian
map). Then $\tilde{\Phi}_{\mathrm{CP}}=|c_{0}|\rho
+|c_{1}|\sum_{n=1}^{3}X_{n}\rho X_{n}$, which is a non-trace preserving
version of the well known independent-errors CP bit-flip map. The code is $%
\mathcal{C}=\mathrm{span}\{|0_{L}\rangle \equiv |000\rangle ,|1_{L}\rangle
\equiv |111\rangle \}$, and $P=|0_{L}\rangle \langle 0_{L}|+|1_{L}\rangle
\langle 1_{L}|$, which satisfies Eq.~(\ref{eq:QEC-H}) with $F_{1}=\sqrt{%
|c_{0}|}I$ and $F_{2,3,4}=\sqrt{|c_{1}|}X_{1,2,3}$. Then by Corollary \ref%
{cor:HM} the same code (and corresponding CP recovery map) also corrects $%
\Phi _{\mathrm{IPF}}$. The CP recovery map $\mathcal{R}_{\mathrm{CP}}$ has
operation elements $R_{0}=P$ and $\{R_{n}=\frac{1}{\sqrt{3}}%
PX_{n}\}_{n=1}^{3}$; indeed, it is easily checked that $\mathcal{R}_{\mathrm{%
CP}}[\Phi _{\mathrm{IPF}}(P\rho P)]=P\rho P$ for any state $\rho \in 
\mathcal{C}$.

\subsubsection{Hermitian recovery maps}

Since Hermitian maps are the most general physical maps, it is natural to
consider Hermitian recovery of Hermitian noise maps. We thus define
\textquotedblleft Hermitian recovery maps\textquotedblright\ $\{\mathcal{R}_{%
\mathrm{H}}\}$ as those Hermitian maps that correct a Hermitian noise map $%
\Phi _{\mathrm{H}}$, i.e., $\mathcal{R}_{\mathrm{H}}\circ \Phi _{\mathrm{H}%
}(\rho )\varpropto \rho $. The following result presents a possible set of
Hermitian recovery maps.

\begin{mycorollary}
\label{cor:Hrec} Consider a Hermitian noise map $\Phi _{\mathrm{H}}(\rho )$ $%
={\sum_{i=1}^{N}}c_{i}K_{i}\rho K_{i}^{\dagger }$ with error operators $%
\{K_{i}\}$ satisfying the relations $PK_{i}^{\dag }K_{j}P=\alpha _{ij}P$.
Any Hermitian map $\mathcal{R}_{\mathrm{H}}(\rho)=\sum_{k}h_{k}R_{k}\rho
R_{k}^{\dag }$ with recovery operators $\{R_{k}\}$ as in Eq. (\ref%
{eq:recovery}) and $\{h_{k}\}\in \mathbb{R}$ corrects the noise map $\Phi _{%
\mathrm{H}}$.
\end{mycorollary}

The proof is given in Appendix~\ref{app:Hrec}, and employs a method similar
to that of the proof of Theorem~\ref{th:suff}.

\begin{figure}[tph]
\centering
\includegraphics[width=3.5in]{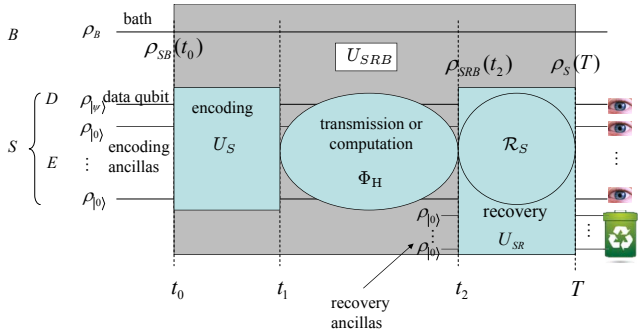} \vspace{-2.5cm}
\caption{The initial system-bath state is the generically non-VQD\ state $%
\protect\rho _{SB}(t_{0})$. The encoded system $S=D+E$ consists of data
qubits $D$ and encoding ancillas $E$. We also include the recovery ancillas $%
R$, which are assumed to be completely isolated until they are brought into
contact with $S$ and $B$ at a later time. Thus the full initial state is $%
\protect\rho _{SB}(t_{0})\otimes \protect\rho _{R}(t_{0})$. The overall
evolution is governed by the unitary $U_{SRB}$ which acts on the system $S$,
the bath $B$, and eventually the recovery ancillas $R$, and is denoted by
the large grey box. The state of the data qubits is $\protect\rho _{|\protect%
\psi \rangle }=\mathrm{Tr}_{E,B}[\protect\rho _{SB}(t_{0})]$, a state which
is as close as possible (by isolating the system) to the desired pure data
state $|\protect\psi \rangle $. The state of each of the encoding ancillas
is $\protect\rho _{|0\rangle }=\mathrm{Tr}_{E^{\prime },B}[\protect\rho %
_{SB}(t_{0})]$, a state which is as close as possible (again, by isolating
the system) to the desired pure encoding ancilla state $|0\rangle $. Here $%
\mathrm{Tr}_{E,B}$ denotes a partial trace over all encoding ancillas and
the bath, $\mathrm{Tr}_{E^{\prime },B}$ denotes a partial trace over all but
one of the encoding ancillas, and the bath. Ideally, the encoding unitary $%
U_{S}$ is then applied to the encoded system. This is of course an
idealization since in reality the encoding operation will not be a perfect
unitary; instead what is really applied is $U_{SRB}(t_{1},t_{0})$, which is
supposedly close to the ideal $U_{S}\otimes I_{R}\otimes I_{B}$. Thus, after
the encoding the total state is $\protect\rho %
_{SRB}(t_{1})=U_{SRB}(t_{1},t_{0})[\protect\rho _{SB}(t_{0})\otimes \protect%
\rho _{R}(t_{0})]U_{SRB}^{\dag }(t_{1},t_{0})$ and the encoded system state
is $\protect\rho _{S}(t_{1})=\mathrm{Tr}_{R,B}[\protect\rho _{SRB}(t_{1})]$.
The system is then passed through the noise channel for the purpose of
either computation or communication, i.e., $\protect\rho %
_{SRB}(t_{2})=U_{SRB}(t_{2},t_{1})\protect\rho _{SRB}(t_{1})U_{SRB}^{\dag
}(t_{2},t_{1})$, whence $\protect\rho _{S}(t_{2})=\mathrm{Tr}_{R,B}[\protect%
\rho _{SRB}(t_{2})]=\Phi _{\mathrm{H}}[\protect\rho _{S}(t_{1})]$, where $%
\Phi _{\mathrm{H}}$ is a Hermitian noise map since $\protect\rho %
_{SRB}(t_{1})$ is generically a non-VQD state due to the initial
non-classical correlations between $S$ and $B$. The goal of the error
correction procedure is to recover the original encoded system state from $%
\protect\rho _{S}(t_{2})$, and to this end we introduce recovery ancillas $R$
at $t_{2}$. Similarly to the encoding ancillas, these recovery ancillas are
each in the state $\protect\rho _{|0\rangle }=\mathrm{Tr}_{S,R^{\prime },B}[%
\protect\rho _{SRB}(t_{2})]$, a state which is as close as possible to the
desired pure recovery ancilla state $|0\rangle $. Next, ideally the recovery
unitary $U_{SR}\otimes I_{B}$ is applied. In reality what is applied is $%
U_{SRB}(T,t_{2})$, which is supposedly close to the ideal $U_{SR}\otimes
I_{B}$. Then the recovery ancillas are discarded and possibly recycled,
leaving the encoded system in the final state $\protect\rho _{S}(T)=\mathrm{%
Tr}_{R,B}[\protect\rho _{SRB}(T)]=\mathcal{R}_{S}[\protect\rho _{S}(t_{2})]$%
, which can be measured. Since $\protect\rho _{SRB}(t_{2})$ is generically
not a VQD state (due to non-classical correlations between $S$ and $R$,
mediated by their mutual interaction with $B$), it is clear that the
recovery map $\mathcal{R}_{S}$ is generically a non-CP Hermitian map. We
recover the CP\ recovery map scenario if, for example, $\protect\rho %
_{SRB}(t_{2})=\protect\rho _{SB}(t_{2})\otimes \protect\rho _{R}(t_{2})$.
The assumption that this is not the case is consistent with the working
premise of this paper and is equivalent in that regard to the assumption
that the initial system-bath state is not of the form $\protect\rho %
_{SB}(t_{0})=\protect\rho _{S}(t_{0})\otimes \protect\rho _{B}(t_{0})$.}
\label{fig}
\end{figure}

\subsubsection{How does non-CP, Hermitian recovery arise?}

In standard QEC theory the recovery map is considered CP. The reason for
this is that the recovery ancillas are introduced after the action of the
noise channel so that they enter in a tensor product state with the encoded
qubits that underwent the noise channel. The recovery map is obtained in the
standard setting by first applying a unitary over the encoded qubits plus
recovery ancillas, then tracing out the recovery ancillas. This is
manifestly a CP\ map over the encoded qubits.

Since we know that the recovery map experienced by the encoded qubits is CP
if and only if the initial state of the encoded and recovery ancilla qubits
has vanishing quantum discord \cite{ShabaniLidar:08}, it is clear how a
non-CP recovery map can be implemented: the recovery ancillas should have
non-vanishing quantum discord with the encoded qubits. Since this will still
be a QDP, the resulting recovery map will be Hermitian according to Theorem %
\ref{linear Rep}.

Such a situation can come about in various ways. For example, a scenario
which is particularly relevant for quantum computation and communication, is
one where the environment causes the recovery ancillas to become
non-classically correlated with the encoded qubits before the recovery
operation can be applied. This is a reasonable scenario since, while the
recovery ancillas are presumably kept pure and isolated from the environment
for as long as possible, at some point they must be brought into contact
with the encoded qubits, and at this point all qubits (encoded and recovery
ancillas) are susceptible to correlations mediated by the environment. This
is shown in Fig. \ref{fig}.

\section{Conclusions}

\label{sec:conc}

This work aimed to fill two gaps: one in the theory of open quantum systems,
and a resulting gap in the theory of quantum error correction. The first gap
had to do with the type of maps that describe open systems given \emph{%
arbitrary} initial states of the total system. In fact, it was not a priori
clear that there should even be a linear map connecting the initial to the
final open system state for arbitrary initial total system states. Building
upon the class of \textquotedblleft special linear states\textquotedblright\
we introduced in \cite{ShabaniLidar:08} we showed here that in fact such a
linear map description does always exist, and moreover, for quantum dynamics
the map is always Hermitian. The map reduces to the completely positive type
if and only if the initial total system state has vanishing quantum discord 
\cite{ShabaniLidar:08}; in all other cases it is Hermitian but not CP. This
result, we argued, impacts the theory of quantum error correction, where
previously the assumption of CP\ maps was taken for granted. In the second
part of this work we filled this gap in QEC theory, by developing a theory
of Linear Quantum Error Correction (LQEC), which generalizes the
CP-map-based standard theory of QEC. We showed that to every linear map $%
\Phi _{\mathrm{L}}$ is associated a CP map which, if correctable, also
provides an encoding with corresponding CP recovery map for $\Phi _{\mathrm{L%
}}$ (Theorem \ref{th:CP-rec}). Moreover, it is possible to find a non-CP
recovery for $\Phi _{\mathrm{L}}$ within a larger class of codes (Theorem %
\ref{th:suff}). From a physical standpoint this result is actually too
general, since only Hermitian maps ever arise from quantum dynamics [to the
extent that the standard quantum dynamical process (\ref{dynamics1}) is
valid]. Hence we specialized LQEC\ to the Hermitian maps case, and showed
that in this case standard QEC theory for CP\ maps already suffices, in the
sense that it is legitimate to replace a given Hermitian noise map by a
corresponding CP\ map obtained simply by taking the absolute values of all
the Hermitian map coefficients. Any QEC code which corrects this CP\ map
will also correct the original Hermitian map (Corollary \ref{cor:HM}).
Nevertheless, there is room for a genuine generalization when one considers
Hermitian maps, since it is also possible to perform QEC using Hermitian
recovery maps (Corollary \ref{cor:Hrec}). We argued that, in fact, recovery
maps will generically be non-CP Hermitian maps, since recovery ancillas that
are introduced into a quantum circuit prior to the recovery step will become
non-classically correlated with the environment and consequently with the
rest of the system.

An interesting open question for future studies is whether the results
presented here have an impact on the threshold for fault tolerant quantum
error correction. For example, note that while CP\ recovery perfectly
returns the encoded state [Eqs. (\ref{eq:RL_CP}) and (\ref{eq:RH_CP})],
non-CP\ recovery only does so up to a proportionality factor which depends
on the details of the noise and recovery maps [$F_{\mathrm{L}}$ in Eq. (\ref%
{eq:F_L}) and $F_{\mathrm{H}}$ in Eq. (\ref{eq:F_H})]. This proportionality
factor -- assuming non-CP\ recovery is applied -- may differ for different
terms in the fault path decomposition \cite{Aharonov:08}, an effect which
may propagate into the value of the fault tolerance threshold. This requires
careful analysis, which is beyond the scope of this paper.

\begin{acknowledgments}
Funded by the National Science Foundation under Grants No. CCF-0726439,
PHY-0802678, and PHY-0803304, and by the United States Department of Defense
(to D.A.L.). Part of this work was done while D.A.L. enjoyed the generous
hospitality of the Institute for Quantum Information at the California
Institute of Technology.
\end{acknowledgments}

\appendix

\section{Proof of Theorem 1}

\label{app:th1}

We use a method similar to Choi's proof for a CP map representation \cite%
{Choi:75}, recently clearly reviewed in Ref. \cite{Leung:03}. The main
difference between the proofs in Refs. \cite{Choi:75,Leung:03} and our proof
is that in the previous proofs positivity allowed for the use of standard
diagonalization, whereas in the absence of positivity we use the singular
value decomposition \cite{Horn:book}.

\begin{proof}
Eq.~(\ref{eq:LM}) immediately implies that $\Phi _{\mathrm{L}}$ is a linear
map. For the other direction, let $\widetilde{M}=\sum_{i,j=1}^{n}|i\rangle
\langle j|\otimes |i\rangle \langle j|=n|\phi \rangle \langle \phi |$, where 
$|i\rangle $ is a column vector with $1$ at position $i$ and $0$'s
elsewhere, and $|\phi \rangle =n^{-1/2}\sum_{i}|i\rangle \otimes |i\rangle $
is a maximally entangled state over $\mathcal{H}\otimes \mathcal{H}$, where $%
\mathcal{H}$ is the Hilbert space spanned by $\{|i\rangle \}_{i=1}^{n}$. $%
\widetilde{M}$ is also an $n\times n$ array of $n\times n$ matrices, whose $%
(i,j)$th block is $|i\rangle \langle j|$. Construct two equivalent
expressions for $(\mathcal{I}\otimes \Phi _{\mathrm{L}})[\widetilde{M}]$,
where $\mathcal{I}$ is the $(n\times n)\times (n\times n)$ identity matrix.
(i) $(\mathcal{I}\otimes \Phi _{\mathrm{L}})[\widetilde{M}]$ is an $n\times
n $ array of $m\times m$ matrices, whose $(i,j)$th block is $\Phi _{\mathrm{L%
}}[|i\rangle \langle j|]$. (ii) Consider a singular value decomposition: $(%
\mathcal{I}\otimes \Phi )[\widetilde{M}]=UDV=\sum_{\alpha }\lambda _{\alpha
}U|\alpha \rangle \langle \alpha |V=\sum_{\alpha }\lambda _{\alpha
}|u_{\alpha }\rangle \langle v_{\alpha }|$. Here $U$ and $V$ are unitary, $D=%
\mathrm{diag}(\{\lambda _{\alpha }\})$ is diagonal and $\lambda _{\alpha
}\geq 0$ are the singular values of $(\mathcal{I}\otimes \Phi _{\mathrm{L}})[%
\widetilde{M}]$. Divide the column (row) vector $|u_{\alpha }\rangle $ ($%
\langle v_{\alpha }|$) into $n$\ segments each of length $m$ and define an $%
m\times n$ ($n\times m$) matrix $E_{\alpha }$ ($E_{\alpha }^{\prime }$)
whose $i$th column (row) is the $i$th segment; then $E_{\alpha }|i\rangle $ (%
$\langle i|E_{\alpha }^{\prime \dagger }$) is the $i$th segment of $%
|u_{\alpha }\rangle $ ($\langle v_{\alpha }|$). Therefore the $(i,j)$th
block of $|u_{\alpha }\rangle \langle v_{\alpha }|$ becomes $E_{\alpha
}|i\rangle \langle j|E_{\alpha }^{\prime \dagger }$.

Equating the two expressions in (i) and (ii) for the $(i,j)$th block of $(%
\mathcal{I}\otimes \Phi _{\mathrm{L}})[\widetilde{M}]$, we find $\Phi _{%
\mathrm{L}}[|i\rangle \langle j|]={\sum_{\alpha }}\lambda _{\alpha
}E_{\alpha }|i\rangle \langle j|E_{\alpha }^{\prime \dagger }$. Since $%
\lambda _{\alpha }\geq 0$ we can redefine $E_{\alpha }$ as $\sqrt{\lambda
_{\alpha }}E_{\alpha }$ and $E_{\alpha }^{\prime }$ as $\sqrt{\lambda
_{\alpha }}E_{\alpha }^{\prime }$, which we do from now on. Finally, the
linearity assumption on $\Phi _{\mathrm{L}}$, together with the fact that
the set $\{|i\rangle \langle j|\}_{i,j=1}^{n}$ spans $\mathfrak{M}_{n}$,
implies Eq.~(\ref{eq:LM}).

Next let us prove Eq. (\ref{eq:QM}) for Hermitian maps. For an old proof
that uses very different techniques see Ref. \cite{Hill:73}. Eq.~(\ref{eq:QM}%
) immediately implies that $\Phi _{\mathrm{H}}$ is a Hermitian map. For the
other direction, associate a matrix $L_{\Phi _{\mathrm{H}}}$ with the
Hermitian map $\Phi _{\mathrm{H}}$:\ $\rho ^{\prime }=\Phi _{\mathrm{H}%
}(\rho )$ $\Longleftrightarrow $ $\rho _{m\mu }^{\prime }=L_{n\nu }^{m\mu
}\rho _{n\nu }$ (summation over repeated indices is implied). Hermiticity of 
$\rho $ and its image $\rho ^{\prime }$ implies $\rho _{\mu m}^{\prime
}=\rho _{m\mu }^{\prime \ast }=L_{n\nu }^{m\mu \ast }\rho _{n\nu }^{\ast
}=L_{n\nu }^{m\mu \ast }\rho _{\nu n}$, i.e., $L_{n\nu }^{m\mu \ast }=L_{\nu
n}^{\mu m}$ \cite{Zyczkowski:04}. We can use this property of $L_{\Phi _{%
\mathrm{H}}}$ to show that if $\Phi _{\mathrm{H}}$ is a Hermitian map, then $%
\mathcal{I}\otimes \Phi _{\mathrm{H}}$ is Hermiticity preserving. Consider $%
\mathcal{M}=\mathcal{M}_{k\xi }^{n\nu }|k\rangle \langle \xi |\otimes
|n\rangle \langle \nu |$. Then $\mathcal{M}^{\prime }=(\mathcal{I}\otimes
\Phi _{\mathrm{H}})[\mathcal{M}]=\mathcal{M}_{k\xi }^{n\nu }|k\rangle
\langle \xi |\otimes \Phi _{\mathrm{H}}(|n\rangle \langle \nu |)=\mathcal{M}%
_{k\xi }^{m\mu }|k\rangle \langle \xi |\otimes L_{n\nu }^{m\mu }|n\rangle
\langle \nu |$. Assume that $\mathcal{M}_{k\xi }^{m\mu \ast }=\mathcal{M}%
_{\xi k}^{\mu m}$. This property holds for $\mathcal{M}=\widetilde{M}=|\phi
\rangle \langle \phi |$ where $|\phi \rangle =\dim (\mathcal{H}%
)^{-1/2}\sum_{i}|i\rangle \otimes |i\rangle $ is a maximally entangled state
over $\mathcal{H}\otimes \mathcal{H}$ ($\mathcal{M}_{\xi k}^{\mu m}\equiv 1$%
). Then $\mathcal{M}^{\prime \dag }=\mathcal{M}_{k\xi }^{m\mu \ast }|\xi
\rangle \langle k|\otimes L_{n\nu }^{m\mu \ast }|\nu \rangle \langle n|=%
\mathcal{M}_{\xi k}^{\mu m}|\xi \rangle \langle k|\otimes L_{\nu n}^{\mu
m}|\nu \rangle \langle n|=\mathcal{M}^{\prime }$. Therefore $(\mathcal{I}%
\otimes \Phi _{\mathrm{H}})[|\phi \rangle \langle \phi |]$ is Hermitian, and
in particular invertible. It follows that the SVD\ used in the proof of
Theorem~\ref{th1} can be replaced by standard diagonalization ($U=V^{\dag }$%
). In this case the left and right singular vectors $|u_{\alpha }\rangle
=\langle v_{\alpha }|^{\dag }$ are the eigenvectors of $(\mathcal{I}\otimes
\Phi _{\mathrm{H}})[|\phi \rangle \langle \phi |]$ and $c_{\alpha }=\lambda
_{\alpha }$ are its eigenvalues. Then $E_{\alpha }=E_{\alpha }^{\prime }$ in
Eq.~(\ref{eq:LM}) and $c_{\alpha }\in \mathbb{R}$.
\end{proof}

We note that by splitting the spectrum of $(\mathcal{I}\otimes \Phi _{%
\mathrm{H}})[|\phi \rangle \langle \phi |]$ into positive and negative
eigenvalues, $\{c_{\alpha }^{+}\geq 0\}$ and $\{c_{\alpha }^{-}\leq 0\}$, we
have as an immediate corollary a fact that was also noted in \cite{Jordan:04}%
: Any Hermitian map can be represented as the difference of two CP\ maps: $%
\Phi (\rho )$ $={\sum_{\alpha }}c_{\alpha }^{+}E_{\alpha }^{+}\rho E_{\alpha
}^{+\dagger }-{\sum_{\alpha }}|c_{\alpha }^{-}|E_{\alpha }^{-}\rho E_{\alpha
}^{-\dagger }$.

\section{Direct Proof of Corollary 1}

\label{app:direct}

\begin{proof}
The operation elements of $\tilde{\Phi}_{\mathrm{CP}}$ are $\{{F}_{i}=\sqrt{{%
|}c_{i}|}K_{i}\}_{i=1}^{N}$, whence $\tilde{\Phi}_{\mathrm{CP}}(\rho )={%
\sum_{i=1}^{N}}F_{i}\rho F_{i}^{\dagger }$. The standard quantum error
conditions (\ref{eq:QEC-CP}) for $\tilde{\Phi}_{\mathrm{CP}}$ is a set of
conditions in terms of the $F_{i}$: 
\begin{equation}
PF_{i}^{\dag }F_{j}P=\beta _{ij}P,\quad i,j\in \{1,\ldots ,N\}.
\label{eq:QEC-H}
\end{equation}%
The existence of a projector $P$ which satisfies Eq.~(\ref{eq:QEC-H}) is
equivalent to the existence of a QEC code for $\tilde{\Phi}_{\mathrm{CP}}$.
Assuming that a code $\mathcal{C}$ has been found (i.e., $P\mathcal{C}=%
\mathcal{C}$) for $\tilde{\Phi}_{\mathrm{CP}}$, we use this as a code for $%
\Phi _{\mathrm{H}}$ and show that the corresponding CP recovery map $%
\mathcal{R}_{\mathrm{CP}}$ is also a recovery map for $\Phi _{\mathrm{H}}$.
Indeed, let $G_{j}\equiv \sum_{i=1}^{N}u_{ij}F_{i}$ be new operation
elements for $\tilde{\Phi}_{\mathrm{CP}}$, i.e., $\tilde{\Phi}_{\mathrm{CP}}=%
{\sum_{j=1}^{N}}G_{j}\rho G_{j}^{\dagger }$, where $u$ is the unitary matrix
that diagonalizes the Hermitian matrix $\beta =[\beta _{ij}]$, i.e., $%
u^{\dag }\beta u=d$. Let $\mathcal{R}_{\mathrm{CP}}=\{R_{k}\}$ be the CP
recovery map for $\tilde{\Phi}_{\mathrm{CP}}$. Assume that $\rho $ is in the
code space, i.e., $P\rho P=\rho $. We now show that $\mathcal{R}_{\mathrm{CP}%
}[\Phi _{\mathrm{H}}(\rho )]=\rho $, i.e., we have CP recovery. First, 
\begin{eqnarray}
\mathcal{R}_{\mathrm{CP}}[\Phi _{\mathrm{H}}(\rho )] &=&\sum_{k}R_{k}\left(
\sum_{i=1}^{N}\frac{c_{i}}{|c_{i}|}F_{i}\rho F_{i}^{\dag }\right)
R_{k}^{\dag }  \notag \\
&=&\sum_{i=1}^{N}\frac{c_{i}}{|c_{i}|}\sum_{j,j^{\prime }=1}^{N}u_{ij}^{\ast
}u_{ij^{\prime }}  \notag \\
&&\times \sum_{k}\left( R_{k}G_{j}P\right) \rho \left( PG_{j^{\prime
}}^{\dag }R_{k}^{\dag }\right) .
\end{eqnarray}%
Now, note that, using Eq. (\ref{eq:QEC-H}): 
\begin{eqnarray}
PG_{k}^{\dag }G_{l}P &=&\sum_{ij}u_{ik}^{\ast }u_{jl}PF_{i}^{\dag
}F_{j}P=\sum_{ij}u_{ik}^{\ast }\beta _{ij}u_{jl}P  \notag \\
&=&d_{k}\delta _{kl}P.
\end{eqnarray}%
Then the polar decomposition yields $G_{k}P=U_{k}(PG_{k}^{\dag
}G_{k}P)^{1/2}=\sqrt{d_{k}}U_{k}P$. The recovery operation elements are
given by 
\begin{equation}
R_{k}=U_{k}^{\dag }P_{k};\quad P_{k}=U_{k}PU_{k}^{\dag }.
\end{equation}%
Therefore $P_{k}=G_{k}PU_{k}^{\dag }/\sqrt{d_{k}}$. This allows us to
calculate the action of the $k$th recovery operator on the $l$th error: 
\begin{eqnarray}
R_{k}G_{l}P &=&U_{k}^{\dag }P_{k}^{\dag }G_{l}P=U_{k}^{\dag
}(U_{k}PG_{k}^{\dag }/\sqrt{d_{k}})G_{l}P  \notag \\
&=&\delta _{kl}\sqrt{d_{k}}P.
\end{eqnarray}%
Therefore, 
\begin{eqnarray}
\mathcal{R}_{\mathrm{CP}}[\Phi _{\mathrm{H}}(\rho )] &=&\sum_{i=1}^{N}\frac{%
c_{i}}{|c_{i}|}\sum_{j,j^{\prime }=1}^{N}u_{ij}^{\ast }u_{ij^{\prime }} 
\notag \\
&&\times \sum_{k}\left( \delta _{kj}\sqrt{d_{k}}P\right) \rho \left( P\sqrt{%
d_{k}}\delta _{kj^{\prime }}\right)  \notag \\
&=&\rho \sum_{i=1}^{N}\frac{c_{i}}{|c_{i}|}\left( udu^{\dag }\right) _{ii} 
\notag \\
&=&(\sum_{i=1}^{N}\frac{c_{i}}{|c_{i}|}\beta _{ii})\rho .  \label{eq:R_CP}
\end{eqnarray}%
Next note that, using condition (\ref{eq:QEC-H}) and trace preservation by $%
\Phi _{\mathrm{H}}$: 
\begin{eqnarray}
PF_{i}^{\dag }F_{i}P &=&\beta _{ii}P\Longrightarrow \sum_{i=1}^{N}\frac{c_{i}%
}{|c_{i}|}\beta _{ii}P  \notag \\
&=&P\sum_{i=1}^{N}\frac{c_{i}}{|c_{i}|}F_{i}^{\dag
}F_{i}P=P\sum_{i=1}^{N}c_{i}K_{i}^{\dag }K_{i}P=P  \notag \\
&\Longrightarrow &\sum_{i=1}^{N}\frac{c_{i}}{|c_{i}|}\beta _{ii}=1.
\end{eqnarray}%
Hence, finally: 
\begin{equation}
\mathcal{R}_{\mathrm{CP}}[\Phi _{\mathrm{L}}(\rho )]=\rho  \label{eq:RH_CP}
\end{equation}%
for any $\rho $ in the codespace.
\end{proof}

\section{Proof of Corollary 2}

\label{app:Hrec}

\begin{proof}
Let $\{{F}_{i}=\sqrt{{|}c_{i}|}K_{i}\}_{i=1}^{N}$; we simply use the
identities given in the proof of the previous theorem -- specifically Eq. (%
\ref{eq:R_CP}) -- to calculate $\mathcal{R}_{\mathrm{H}}\circ \Phi _{\mathrm{%
H}}(\rho )$%
\begin{eqnarray}
\mathcal{R}_{\mathrm{H}}[\Phi _{\mathrm{H}}(\rho )]
&=&\sum_{k}h_{k}R_{k}\left( \sum_{i=1}^{N}\frac{c_{i}}{|c_{i}|}F_{i}\rho
F_{i}^{\dag }\right) R_{k}^{\dag }  \notag \\
&=&\sum_{i=1}^{N}h_{k}\frac{c_{i}}{|c_{i}|}\sum_{j,j^{\prime
}=1}^{N}u_{ij}^{\ast }u_{ij^{\prime }}\times  \notag \\
&&\sum_{k}\left( \delta _{kj}\sqrt{d_{k}}P\right) \rho \left( P\sqrt{d_{k}}%
\delta _{kj^{\prime }}\right)  \notag \\
&=&P\rho P\sum_{i=1}^{N}\frac{c_{i}}{|c_{i}|}\sum_{k}h_{k}u_{ik}^{\ast
}u_{ik}d_{k}  \notag \\
&=&F_{\mathrm{H}}P\rho P\propto \rho ,
\end{eqnarray}%
where 
\begin{equation}
F_{\mathrm{H}}\equiv \sum_{i=1}^{N}\frac{c_{i}}{|c_{i}|}\left( udhu^{\dag
}\right) _{ii},  \label{eq:F_H}
\end{equation}%
where $h\equiv \mathrm{diag}(\{h_{k}\})$, and $F_{\mathrm{H}}$ is a
\textquotedblleft correction factor\textquotedblright\ for Hermitian\
recovery of Hermitian noise maps, which was $1$ in the case of CP\ recovery,
above.
\end{proof}


\end{document}